\documentclass[prd,preprintnumbers,showpacs,
showkeys,superscriptaddress,showpacs,nofootinbib,byrevtex,fleqn]{revtex4}
\usepackage{amsmath,amsfonts,amssymb,amscd,amsxtra,amsthm}
\usepackage{graphicx,wrapfig,color}
\usepackage{bm}

\begin{document}
\preprint{INHA-NTG-03/2012}

\title{Energy-momentum tensor form factors of the nucleon in nuclear
  matter}
\author{Hyun-Chul Kim}
\affiliation{Department of Physics, Inha University,
Incheon 402-751, Republic of Korea}
\affiliation{Department of Physics, University of Connecticut,
  Storrs, CT 06269, U.S.A.}
\affiliation{School of Physics, Korea Institute for Advanced Study,
  Seoul 130-722, Republic of Korea}

\author{Peter Schweitzer}
\affiliation{Department of Physics, University of Connecticut,
  Storrs, CT 06269, U.S.A.}

\author{Ulugbek Yakhshiev}
\affiliation{Department of Physics, Inha University,
Incheon 402-751, Republic of Korea}

\date{September, 2012}
\begin{abstract}
The nucleon form factors of the energy-momentum tensor are studied
in nuclear medium in the framework of the in-medium modified Skyrme model.
We obtain a negative $D$-term, in agreement with results from other
approaches, and find that medium effects make the value of $d_1$
more negative.
\end{abstract}
\pacs{
  12.39.Fe, 
  12.39.Dc, 
  21.65.-f, 
    }


\keywords{Energy momentum tensor, Form factors, Nucleon,
Chiral soliton, Nuclear matter}

\maketitle

\textbf{1.}
Understanding the structure of the nucleon has been one of
fundamental issues well over decades. While the scalar, vector and
axial-vector properties of the nucleon have been studied extensively
and comprehended to a great extent,
its energy-momentum tensor (EMT) form factors have been drawn to
attention only quite recently, long after Pagels proposed
them~\cite{Pagels:1966zza}. The reason can be found in the fact that
it is very difficult to get direct access to these form factors
experimentally. However, the generalized parton distributions
(GPDs)~\cite{Muller:1998fv, Ji:1996nm, Collins:1996fb,
  Radyushkin:1997ki}, which are accessible via hard exclusive
reactions~\cite{Saull:1999kt, Adloff:2001cn, Airapetian:2001yk,
  Stepanyan:2001sm, Ellinghaus:2002bq, Chekanov:2003ya, Aktas:2005ty,
  Airapetian:2006zr, Hall-A:2006hx, Girod:2007aa}, make it possible to
extract information on the EMT form factors of the
nucleon. In particular, certain Mellin moments of the GPDs can be
expressed in terms of the EMT form factors~\cite{Ji:1996nm, Ji:1996ek,
  Polyakov:2002yz}.

The nucleon matrix element of the total symmetric EMT are
parameterized by three form factors as follows
\cite{Ji:1996ek,Polyakov:2002yz}
\begin{equation}
  \langle p^\prime| \hat T_{\mu\nu} (0) |p\rangle
  = \bar u(p^\prime,\,s') \left[M_2(t)\,\frac{P_\mu P_\nu}{M_N}+
  J(t)\ \frac{i(P_{\mu}\sigma_{\nu\rho}+P_{\nu}\sigma_{\mu\rho})
    \Delta^\rho}{2M_N}
  + d_1(t)\,
  \frac{\Delta_\mu\Delta_\nu-g_{\mu\nu}\Delta^2}{5M_N}\right] u(p,\,s)\, ,
  \label{Eq:EMTff}
\end{equation}
where $P=(p+p')/2$, $\Delta=(p'-p)$ and $t=\Delta^2$.
$M_N$ is the nucleon mass, and $u(p,\,s)$ denotes the nucleon spinor
with the polarization vector $s$ defined such that it is given as
$(0,\,{\bm s})$ in the rest frame in which $\bm s$ designates the axis
of the spin quantization.
The recent interest about the EMT form factors was stimulated by
the fact that it is possible to define in QCD and to access in
experiment the separate quark and gluon contributions to the form
factors. The only presently known non-trivial piece of information is
the decomposition of $M_2(t)$ at the zero-momentum transfer, which
reveals that about 1/2 of the momentum of a fast moving nucleon
is carried by quarks, and the other half by gluons.
$J(t)$ provides analogous information on how the total angular
momentum of quarks and gluons makes up the nucleon spin, but this
information is presently not known.
The interpretation of the last form factor $d_1(t)$ in
Eq.(\ref{Eq:EMTff}) is less trivial but of equal significance for
understanding the nucleon structure. It provides information on how
the strong forces are distributed and stabilized in the
nucleon~\cite{Polyakov:2002yz,Polyakov:2002wz}.

In all theoretical approaches where it was studied so far,
$d_1\equiv d_1(0)$ was found negative. This feature is expected to be
deeply rooted in the spontaneous breakdown of chiral symmetry
\cite{Polyakov:1999gs,Kivel:2000fg,Goeke:2001tz}.
The form factor $d_1(t)$ can be extracted from the
beam charge asymmetry in deeply virtual Compton
scattering~\cite{Kivel:2000fg}.
The EMT form factors of the nucleon have been studied
in a variety of theoretical approaches:
in lattice
QCD~\cite{Mathur:1999uf,Hagler:2003jd,Gockeler:2003jf, Negele:2004iu,
Brommel:2007sb, Bratt:2010jn, Liu:2012nz},
in chiral perturbation theory
\cite{Chen:2001pv, Belitsky:2002jp,Ando:2006sk,Diehl:2006ya,Dorati:2007bk},
in the chiral quark-soliton model
\cite{Wakamatsu:2005vk, Wakamatsu:2006dy,
  Wakamatsu:2007uc, Goeke:2007fp, Goeke:2007fq, Petrov:1998kf,
Schweitzer:2002nm, Ossmann:2004bp} as well as in the Skyrme
model~\cite{Cebulla:2007ei}.
Those of nuclei have also been studied
\cite{Polyakov:2002yz, Guzey:2005ba, Liuti:2005gi, Scopetta:2009sn}.

It is also of great importance to understand how the nucleon undergoes
changes in nuclear matter. Studying the EMT form factors of the nucleon
in medium offers a new perspective on the internal structure of the
nucleon, and an important step towards the understanding of how nucleon
properties are modified in nuclei.
The first experimental study of deeply virtual Compton
scattering on (gaseous) nuclear targets (H, He, N, Ne, Kr, Xe)
was reported in \cite{Airapetian:2009bi}. The admittedly sizable
error bars of this first experiment did not allow to observe
nuclear modifications. More precise future experiments of this
type can in principle provide information on nuclear modifications
of EMT form factors.

Thus, in the present Letter, we aim at investigating the
form factors of the total EMT of the nucleon in nuclear matter within
the framework of an in-medium modified SU(2) Skyrme model, extending
the previous work~\cite{Cebulla:2007ei}. The Skyrme
model~\cite{Skyrme:1961vq,Adkins:1983ya} provides a simple framework
for the nucleon and connects chiral dynamics to the baryonic sector
explicitly. Hence, the model can be easily extended to nuclear matter,
modifications of chiral properties of the pion being taken into
account. The changes of the nucleon in nuclear matter have been
already examined within the in-medium modified Skyrme
model~\cite{Rakhimov:1996vq, Rakhimov:1998hu}. In
Ref.~\cite{Yakhshiev:2010kf}, the model has been further elaborated,
the stabilizing term being refined in medium, which we will take 
as our framework to study the EMT form factors of the nucleon in nuclear
matter.

\vspace{0.8cm}
\textbf{2.}
We start with the in-medium modified chiral
Lagrangian~\cite{Yakhshiev:2010kf}\footnote{From now on,
the asterisks in expressions indicate the medium modified quantities
which depend on the medium-dependent functions
explicitly. Otherwise, we use the same symbol without any asterisk.} 
\begin{equation}
  \label{Eq:Lagrangian}
{\mathcal L}^*=
        \frac{F_\pi^2}{16}\;{\rm Tr}(\partial_0 U\partial^0 U^\dag)
-\alpha_p\frac{F_\pi^2}{16}\;{\rm Tr}({\bm\nabla} U)\cdot({\bm\nabla} U^\dag)
    +  \frac{1}{32e^2\gamma} \;{\rm Tr}
    [U^\dag(\partial_\mu U),U^\dag(\partial_\nu U)]^2
    +  \alpha_s\frac{m_\pi^2F_\pi^2}{8}\;{\rm Tr}(U-2)\,,
\end{equation}
where $U$ denotes the SU(2) pion field,  $F_\pi$ the pion decay constant,
$e$ a dimensionless parameter, and $m_\pi$ the pion mass.
The medium modifications are contained in the following functions 
\cite{Rakhimov:1996vq, Rakhimov:1998hu, Yakhshiev:2010kf, Ericsonbook}
  \begin{eqnarray}
   \alpha_p(\rho) &=& 1-\chi_p(\rho)\, , \;\;\;
    \chi_p(\rho) = \frac{4\pi c_0 \rho}{\eta+4\pi c_0 g^\prime \rho}\;,\;\;\;
    \eta = 1+\frac{m_\pi}{M_N}
\label{Eq:alpha_p}
    \\
    \alpha_s(\rho) &=& 1+\frac{\chi_s(\rho)}{m_\pi^2}\, , \;\;\;
    \chi_s(\rho) = - 4\pi\eta b_0 \rho \;,\label{Eq:alpha_s} \\
    \gamma(\rho)&=&\exp\left(-\frac{\gamma_{\rm num}\rho}{1+\gamma_{\rm
      den}\rho}\right)\,.
\label{Eq:gamma}
  \end{eqnarray}
The $\alpha_{s,p}$ depend on the $S$- and $P$-wave pion-nucleus
scattering lengths, volumes ($b_0$ and $c_0$), and the density $\rho$
of nuclear matter \cite{Rakhimov:1996vq, Ericsonbook}, and
$g^\prime$ is the Lorentz-Lorenz or correlation parameter. 
Near the threshold, the energy dependence of $b_0$ and $c_0$ being
ignored and nuclear matter being approximated to be homogeneous,
$\alpha_{s,p}$ simply become functions of the given 
nuclear density. Similarly, the function $\gamma$ in
Eq.(\ref{Eq:gamma}) describes the modification of the Skyrme term in
nuclear matter proposed in Ref.~\cite{Yakhshiev:2010kf} with
$\gamma_{\mathrm{num}}$ and $\gamma_{\mathrm{den}}$ fitted to the
coefficient of the volume term in the semiempirical mass formula.  
We can treat the modified chiral Lagrangian in terms of the
renormalized effective
constants $F_{\pi,t}^*=F_{\pi,t}=F_\pi$,
$F_{\pi,s}^*=\alpha_p^{1/2}F_\pi$, $e^*=\gamma^{1/2}e$, and
$m_\pi^*=(\alpha_s/\alpha_p)^{1/2}m_\pi$.
The behavior of these parameters in nuclear matter is
consistent with those in chiral perturbation theory
\cite{Meissner:2001gz} and QCD sum rules~\cite{Kim:2003tp}.

Homogeneous nuclear matter allows us to keep
the hedgehog Ansatz for the pion field,
i.e. $U=\exp[i\mathbf{\tau}\frac{\bm
  r}r F(r)]$ with a profile function $F(r)$ in contrast to the case of
local-density approximations for finite 
nuclei \cite{Yakhshiev:2001ht,Meissner:2008mr}.
Consequently, one can immediately write the classical mass functional
as
\begin{equation}
 \label{Eq:Esol}
    M_{\rm sol}^*[F]
  = 4\pi \int\limits_0^\infty d r\; r^2\left[
    \frac{F_{\pi,s}^{*2}}{8}\left(\frac{2\sin ^2 F}{r^2}+F^{\prime2}\right)
  + \frac{\sin^2 F}{2e^{*2}\,r^2}\left(\frac{\sin^2
      F}{r^2}+2F^{\prime2}\right)
  + \frac{m_\pi^{*2} F_{\pi,s}^{*2}}{4}\,(1-\cos F) \right]\,,
\end{equation}
where $F'=dF/dr$.
The minimization of the mass functional~(\ref{Eq:Esol}) leads to the
equation for 
$F(r)$ as follows
\begin{equation}
\left(\frac{r^2}{4}+\frac{2F\,\sin^2F}{
  e^{*2}F_{\pi,s}^{*2}}\right)F^{\prime\prime}
    +\frac{rF^\prime}{2}
    +\frac{F^{\prime2}\,\sin 2F}{e^{*2}F_{\pi,s}^{*2}}
    -\frac{\sin 2F}{4}
    -\frac{\sin^2 F\,\sin 2F}{e^{*2}F_{\pi,s}^{*2} r^2}
    -\frac{m_\pi^{*2}\,r^2 \sin F}{4} = 0\,
\label{Eq:diff-eq}
\end{equation}
with the 
boundary conditions 
$F(0)=\pi$ and $F(r)\to 0$ as $r\to\infty$ 
imposed by the
unit topological number of the chiral soliton.
Having performed a
collective quantization, we arrive at the modified collective
Hamiltonian
\begin{equation}
\label{Eq:H-rot}
    H^* = M_{\rm sol}^* + \frac{{\bm{J}}^2}{2\Theta^*}
          = M_{\rm sol}^* + \frac{{\bm{I}}^2}{2\Theta^*} \,,\;\;\;
    \Theta^* = \frac{2\pi}{3}\int\limits_0^\infty d r\;
    r^2 s^2\biggl[F_{\pi,t}^2+\frac{4 F^{\prime2}}{e^{*2}}
    +\frac{4 s^2}{e^{*2}r^2}\biggr]\;,
\end{equation}
where ${\bm J}^2$ and ${\bm I}^2$ are the squared collective spin and
isospin operators, respectively, which act on the nucleon or
$\Delta$ wave functions obtained from the diagonalization of the
collective Hamiltonian~\cite{Adkins:1983ya}, and $\Theta^*$ is
the moment of inertia of the soliton.
A consistent description of the EMT form factors requires
either to minimize the energy functional including rotational
corrections, or to consider for the nucleon mass and other observables
only the leading contribution in the limit of large number of colors
$N_c$ \cite{Cebulla:2007ei}.
In this work we will follow Ref.~\cite{Cebulla:2007ei} and choose
the second option. In particular, this means that in our treatment
the nucleon and $\Delta$ masses (in vacuum or in medium) are degenerate
and given by the minimum of the mass functional (\ref{Eq:Esol}).

Input parameters in the Skyrmion sector are fixed as
$m_\pi=135\,{\rm MeV}$, $F_\pi = 108.78\,{\rm MeV}$, $e = 4.854$
following Ref.~\cite{Yakhshiev:2010kf}.
Those relevant to nuclear matter are determined by reproducing the
coefficient of the volume term in the semiempirical mass formula
and the experimental data for the compressibility of nuclear
matter~\cite{Yakhshiev:2010kf}:
$b_0=-0.024~m_\pi^{-1}$, $c_0=0.09~m_\pi^{-3}$, $g'=0.7$,
$\gamma_{\rm num}=0.797~m_\pi^{-3}$ and
$\gamma_{\rm den}=0.496~m_\pi^{-3}$.
All observables will be given as functions of $\rho/\rho_0$ with
normal nuclear matter density $\rho_0=0.5m_\pi^3$.

\textbf{3.}
We are now in a position to calculate the EMT form factors of the
nucleon in nuclear matter. Since the details of the general
formalism have already been presented in Ref.~\cite{Cebulla:2007ei}
in free space, we briefly recapitulate only the necessary formulae
here. The components of the static EMT \cite{Polyakov:2002yz}
are given as follows:
\begin{eqnarray}
 T_{00}^*(r)  &=&
        \frac{F_{\pi,s}^{*2}}{8}\biggl(\frac{2\sin^2 F}{r^2}+F^{\prime2}\biggr)
        +\frac{\sin^2 F}{2\,e^{*2}\,r^2}\biggl(\frac{\sin^2 F}{r^2}+2F^{\prime2}
    \biggr)+\frac{m_\pi^{*2} F_{\pi,s}^{*2}}{4}\,(1-\cos F)\,,
    \label{Eq:T00-Skyrme}
\\
T_{0k}^*(\bm{r},\bm{s})  &=&
     \frac{\epsilon^{klm}r^ls^m}{({\bf s}\times{\bf
         r})^2}\;\rho_J^*(r)\,, \cr
    \label{Eq:T0k-Skyrme}
T_{ij}^*(\bm{r}) &=& s^*(r)\left(\frac{r_ir_j}{r^2}-\frac 13\,\delta_{ij}\right)
        + p^*(r)\,\delta_{ij}\, ,
    \label{Eq:Tij-Skyrme}
\end{eqnarray}
where $T_{00}^*$ is called the energy density. The density of the
angular momentum $\rho_J^\ast$, the pressure
density, and the shear force density are expressed respectively as
\begin{eqnarray}
    \label{Eq:angular}
 \rho_J^*(r) &=& \frac{\sin^2 F}{12\,\Theta^*}
        \biggl[F_{\pi,t}^2+\frac{4 }{e^{*2}} F^{\prime2}+\frac{4
          }{ e^{*2}\, r^2}\sin^2 F\biggr] \,,\\
    \label{Eq:pressure}
    p^*(r) &=& -\frac{
      F_{\pi,s}^{*2}}{24} \biggl(\frac{2}{r^2}\sin^2 F + F^{\prime2} \biggr)
    +\frac{\sin^2 F}{6 e^{*2}\,r^2}\biggl(\frac{\sin^2
      F}{r^2}+2F^{\prime2} \biggr) - \frac{
      m_\pi^{*2} F_{\pi,s}^{*2}}{4}\,(1-\cos F) \, , \\
    \label{Eq:shear}
    s^*(r) &=& \biggl(\frac{
      F_{\pi,s}^{*2}}{4}+\frac1{e^{*2}\,r^2}\sin^2 F \biggr)
    \biggl(F^{\prime2}-\frac{1}{r^2}\sin^2 F \biggr)\;.
\end{eqnarray}
The three form factors in Eq.(\ref{Eq:EMTff}) are finally given in the
large $N_c$ limit as
\begin{eqnarray}
M_2^*(t)-\frac{t}{5M_N^{*2}}\,d_1^*(t)
        &=& \frac{1}{M_N^*}\int\mathrm{d}^3
        \bm{r}\;T_{00}^*(r)\;j_0(r\sqrt{-t})\,,
        \label{Eq:M2-d1-model-comp}\\
        d_1^*(t)
        &=& \frac{15 M_N^*}{2}\int\mathrm{d}^3 \bm{r} \;p^*(r)
        \;\frac{j_0(r\sqrt{-t})}{t} \,,
        \label{Eq:d1-model-comp}\\
        J^*(t)
        &=& 3
    \int\mathrm{d}^3\bm{r}\;\rho_J^*(r)\;\frac{j_1(r\sqrt{-t})}{r\sqrt{-t}}\;,
    \label{Eq:J-model-comp}
\end{eqnarray}
where $j_0(z)$ and $j_1(z)$ represent the spherical Bessel functions
of order 0 and 1, respectively. At the zero momentum transfer $t=0$,
$M_2^*(0)$ and $J^*(0)$ are normalized as
\begin{equation}
  \label{eq:norm}
M_2^*(0) \;=\; \frac{1}{M_N^*}\int\mathrm{d}^3 \bm{r}\;T_{00}^*(r) = 1
\,,\;\;\;\;\;
 J^*(0) \;=\; \int\mathrm{d}^3\bm{r}\;\rho_J^*(r)=\frac12\;.
\end{equation}
These relations were proven in \cite{Cebulla:2007ei} for a Skyrmion
in free space, and we have checked that they hold also in nuclear
matter.
It is a feature of the approach with homogeneous nuclear matter
that all expressions and proofs are formally analogous to the free
nucleon case. In our approach, only the parameters of the Skyrme model
are modified, but not the structure of the Lagrangian. 
(Of course, for a free nucleon the EMT is a Lorentz tensor.
In nuclear medium we deal with a conserved Noether current.)
Another important relation for $p^*(r)$ that is a consequence
of the EMT conservation is the stability condition
\begin{eqnarray}
 \int_0^\infty dr\,r^2\, p^*(r)\, \label{eq:stable} = 0 \,.
\end{eqnarray}
Again we verified that the analytic proofs of Eq.~(\ref{eq:stable})
formulated in \cite{Cebulla:2007ei} for a Skyrmion in free space can
be carried over straightforwardly to the medium situation.
Finally the constant $d_1^\ast$ in nuclear matter is given by
\begin{eqnarray}
d_1^* \;=\; 5\pi M_N^* \int_0^\infty dr\,r^4\, p^*(r)
      \;=\; - \frac{4\pi M_N^*}{3} \int_0^\infty d r \,r^4\,
      s^*(r)
\label{eq:d1}
\end{eqnarray}
The conservation of the EMT implies that $p^*(r)$ and $s^*(r)$
are connected to each other by the differential equation
$\frac23\,s^{*\prime}(r)+\frac2r\,s^*(r)+ p^{*\prime}(r)=0$
which is at the origin of the equivalent expressions
for $d_1^\ast$ in (\ref{eq:d1}). We have checked that also
in the medium-modified Skyrme model this differential equation holds, 
and the different expressions in Eq.~(\ref{eq:d1}) yield the same
result for $d_1^\ast$, which is another demonstration
of the theoretical consistency of the approach.
Thus, the shear force density is directly related to the pressure
density, so that we will concentrate only on the pressure here.

In the chiral quark-soliton model \cite{Goeke:2007fp} it was
observed that the negative sign of $d_1$ is a consequence of
stability, which requires $p(r)>0$ (repulsion) in the inner region,
and $p(r)<0$ (attraction) in the outer region, which must balance
each other according to Eq.~(\ref{eq:stable}).
By inserting a factor of $r^2$ in the stability integral
(\ref{eq:stable}) 
one basically obtains the expression for $d_1$ in Eq.~(\ref{eq:d1})
in terms of $p(r)$. The ``additional factor'' $r^2$ gives more
weight to the outer region and is responsible for the negative sign
of $d_1$. This pattern was also observed for the Skyrmion in free
space~\cite{Cebulla:2007ei}. Below we will recover this picture,
with important medium modifications.

\newpage
\textbf{4.}
We now proceed to present the results of the EMT form factors of the
nucleon in nuclear matter and discuss their physical implications.
In Table~\ref{tab:1}, we list the quantities relevant to the EMT
form factors and their densities.
First we want to mention that the energy density in the center of the
nucleon is rather sensitive to the input parameters in the Skyrmion
sector. In fact, the energy density in the center of the nucleon in
free space in this work turns out to be
$T_{00}(0) = 1.45~\mathrm{GeV}\cdot \mathrm{fm}^{-3}$, while
$T_{00}(0) =2.28~\mathrm{GeV}\cdot
\mathrm{fm}^{-3}$ was obtained in Ref.~\cite{Cebulla:2007ei}.  The
reason lies in the different parameter set we use in the present
approach.

As the density of the surrounding nuclear medium increases from zero
to normal nuclear matter density, the energy density in the center of
the nucleon decreases by about a factor of 2.
At the same time we observe that the mean square radius for the energy
density
\begin{equation}
  \label{eq:r00}
    \langle r_{00}^2\rangle^* \;=\;
    \frac{\int \mathrm{d}^3 \bm{r}\,r^2 \,T_{00}^*(r)}{\int
      \mathrm{d}^3 \bm{r} \,T_{00}^*(r)}\;
\end{equation}
increases. This implies that the nucleon is swollen in nuclear matter
and its energy density spreads to larger distance in comparison with
that in free space.

\begin{figure}[h]

\hspace{-2cm} \centerline{
\includegraphics[height=6.45cm]{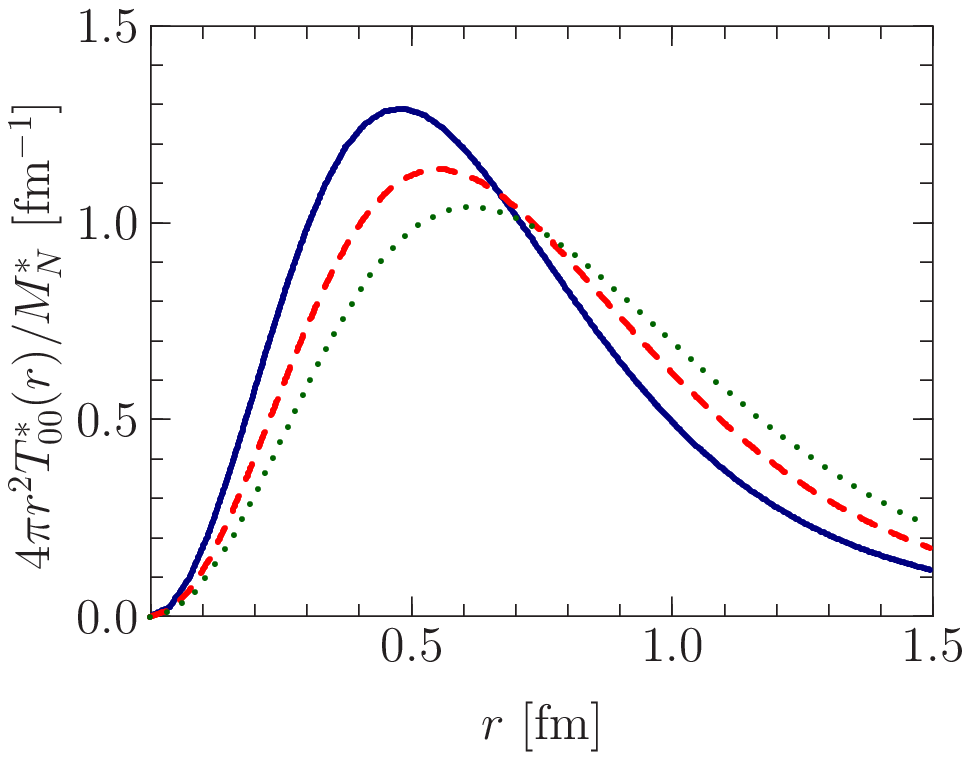}\hspace{0.5cm}
\includegraphics[height=6.45cm]{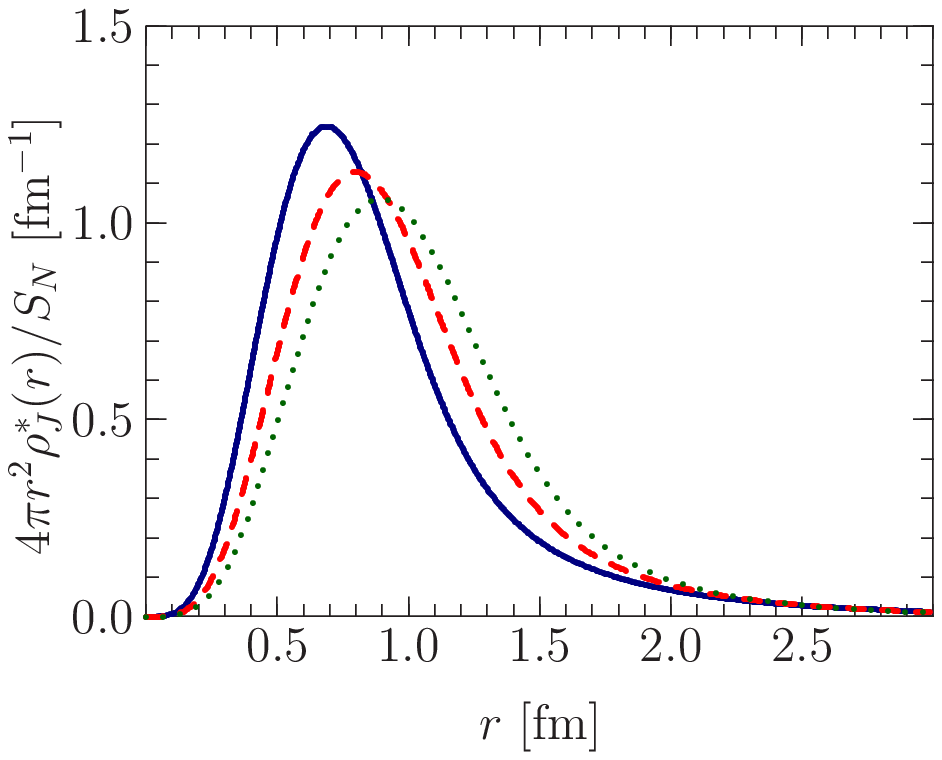}
}
    \caption{(Color online)
In the left panel, the energy densities of the nucleon normalized by the
nucleon mass, $4\pi r^2 T_{00}^* (r)/M_N^*$, are presented as
functions of $r$. The solid curve draws that in free space, whereas
the dashed and dotted ones depict those at the density $\rho=0.5\,\rho_0$
and at normal density $\rho_0$, respectively.
In the right panel, the densities of the angular momentum normalized
by the nucleon spin $S_N=1/2$, $4\pi r^2 \rho_J^* (r)/S_N$ are
rendered as functions of $r$. The other notations are the same as
in the left panel.
}
\label{fig:1}
\end{figure}

Let us more closely look into the change of the energy density in
nuclear matter. The left panel of Fig.~\ref{fig:1} shows how the
energy density $T_{00}^*(r)$ changes in nuclear matter. We present in
fact $4\pi r^2 T_{00}^*(r)$ normalized by the effective in-medium
nucleon mass $M_N^*$
such that the integration of the curve in the
left panel of Fig.~\ref{fig:1} yields unity. The solid curve
draws it in free space, while the dashed and dotted ones depict
those at $\rho=0.5\,\rho_0$ and at normal density ($\rho=\rho_0$),
respectively. Note that the nucleon mass $M_N^*$ also changes as the
density does. As $\rho$ increases, the energy density gets weaker and
shifted to the large distance. Moreover, at a finite $\rho$, it falls
off more slowly than that in free space, as the distance $r$ gets
larger.

%
\begin{table}[h!]
    \caption{
    \label{tab:1}
  Different quantities related to the nucleon EMT densities and their
  form factors: $T_{00}^*(0)$ denotes the energy in the center of the nucleon;
  $\langle r_{00}^2\rangle^*$ and $\langle r_J^2\rangle^*$ depict
  the mean square radii for the energy and angular momentum densities,
  respectively;  $p^*(0)$ represents the pressure in the center of
  the nucleon, whereas $r_0^*$ designates the position where the
  pressure changes its sign; $d_{1}^*$ is the value of the $d_1^*(t)$ form factor
  at the zero momentum transfer.
  }
\vspace{0.2cm}
    \begin{ruledtabular}
    \begin{tabular}{ccccccc}
$\rho/\rho_0$&
$T_{00}^*(0)$ &
$\langle r_{00}^{2}\rangle^*$  &
$\langle r_J^{2}\rangle^*$  &
$p^*(0)$      &
$r_0^*$       &
$d_{1}^*$
\cr
&
$[\mathrm{GeV}\cdot \mathrm{fm}^{-3}]$&
$[\mathrm{fm}^2]$&
$[\mathrm{fm}^2]$&
$[\mathrm{GeV}\cdot \mathrm{fm}^{-3}]$&
$[\mathrm{fm}]$ &
$\phantom{XX}$ \\
    \hline
0    & 1.45 & 0.68& 1.09    & 0.26& 0.71&-3.54 \\
0.5 &0.96  & 0.83& 1.23   & 0.18& 0.82&-4.30 \\
1.0 &0.71  & 0.95& 1.35   & 0.13& 0.90&-4.85 \\
\end{tabular}
\end{ruledtabular}
\end{table}
%
%
\newpage
The density for the angular momentum $\rho_J^*(r)$, which is related
to $T_{0k}^*$, vanishes in the center of the nucleon.
The corresponding mean square radius $\langle r_J^2 \rangle^*$
which is defined analogously to  $\langle r_{00}^2 \rangle^*$ in
Eq.~(\ref{eq:r00}) starts to increase mildly as the density of
nuclear matter increases as shown in Table~\ref{tab:1} and
exhibits similar behavior to $\langle r_{00}^2 \rangle^*$.

Knowing the pressure density is also of great significance, which
contains information on the spatial components of the EMT.
In the fifth column of Table~\ref{tab:1}, we list the values of the
pressure in the center of the nucleon, $p(0)^*$, given nuclear matter
densities. It decreases with $\rho$ increased. In the meanwhile, the
position $r_0^*$, where the pressure changes the sign, becomes larger
when $\rho$ is switched on, which is in line with the features of the
mean square radii discussed previously. That is, one can say that the
shape of the nucleon is bulging out with nuclear matter.

\begin{figure}[t]
\centerline{
\includegraphics[height=6.68cm]{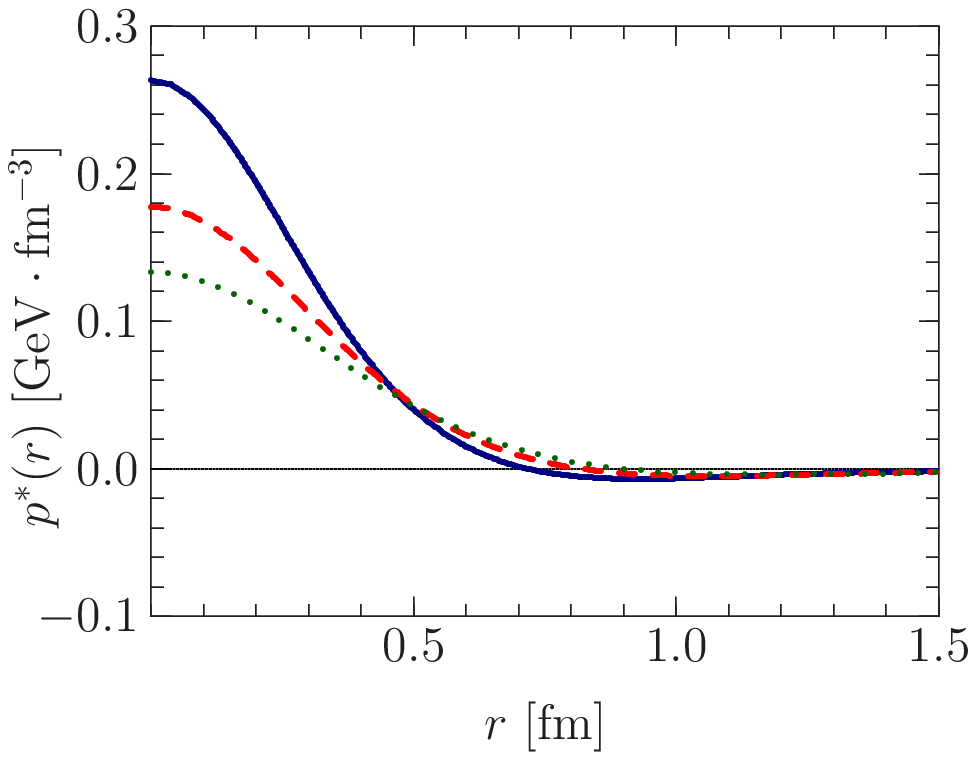} \hspace{0.5cm}
\includegraphics[height=6.68cm]{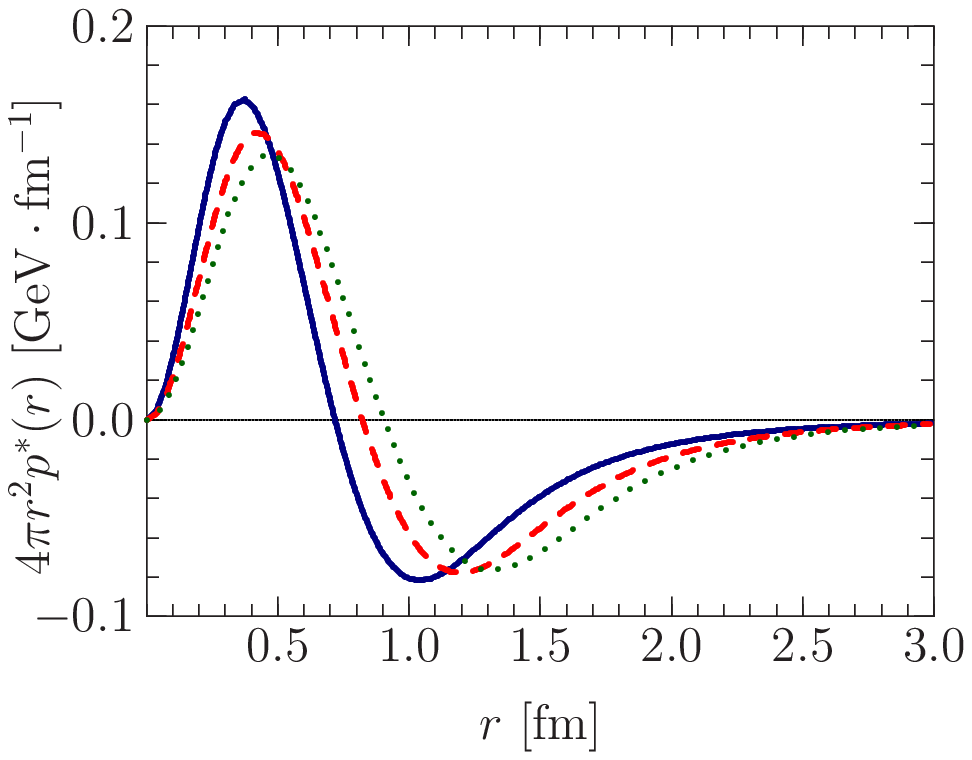}
}
\caption{(Color online)
The pressure densities $p^*(r)$ and $4\pi r^2 p^*(r)$ as functions of
$r$ in the left and right panels, respectively. Notations are the same
as in Fig.~\ref{fig:1}.
}
\label{fig:2}
\end{figure}
Figure~\ref{fig:2} shows how the pressure is distributed in the
nucleon. As discussed already in Ref.~\cite{Cebulla:2007ei}, the
positive sign of the pressure for $r<r_0$ indicates repulsion,
while the negative one in the region $r > r_0$ signifies attraction.
In the Skyrme model, the repulsive part is provided by the
4-derivative stabilizing (Skyrme) term, whereas the
attractive one mainly comes from the kinetic term. Since 
the coefficient $e$ is related to the vector meson
coupling~\cite{Ecker:1988te}, such a distribution of the pressure can
be interpreted as follows: While its repulsive part (or core part) is
mainly governed by the vector mesons ($\rho$ meson), the attractive
part (or long-range part) is explained solely by the pions. This
picture
\begin{figure}[h!]
\centerline{\hspace{-1.5cm}
\includegraphics[height=6.5cm]{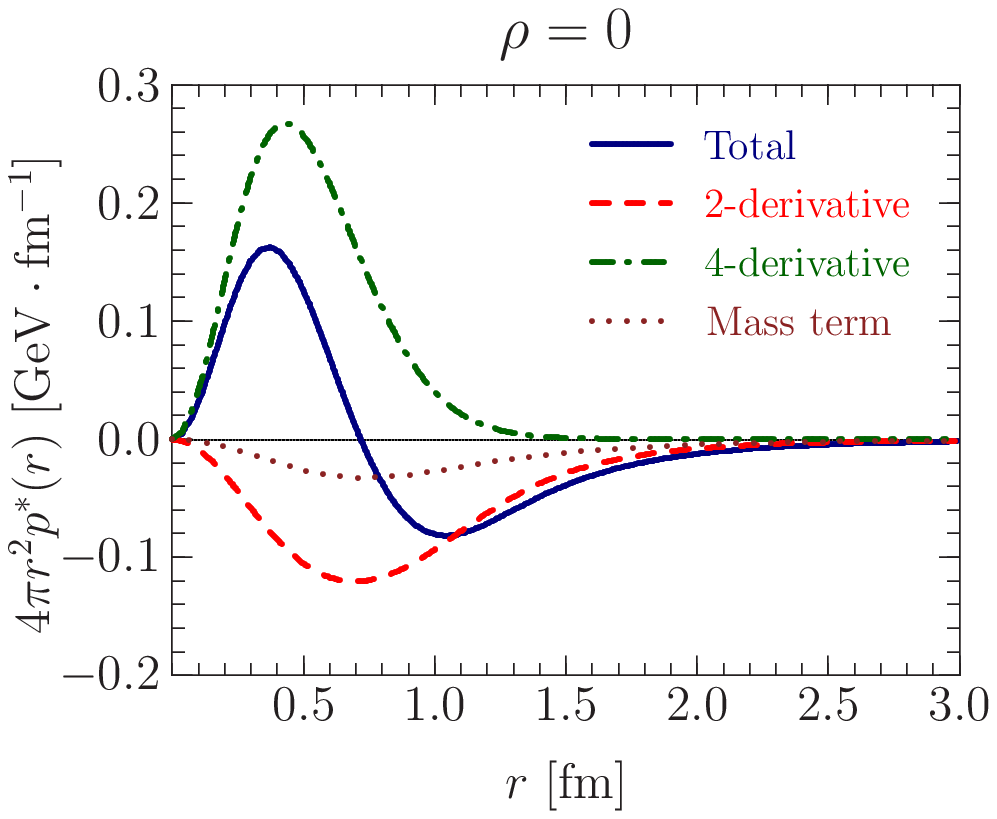} \hspace{-0.8cm}
\includegraphics[height=6.5cm]{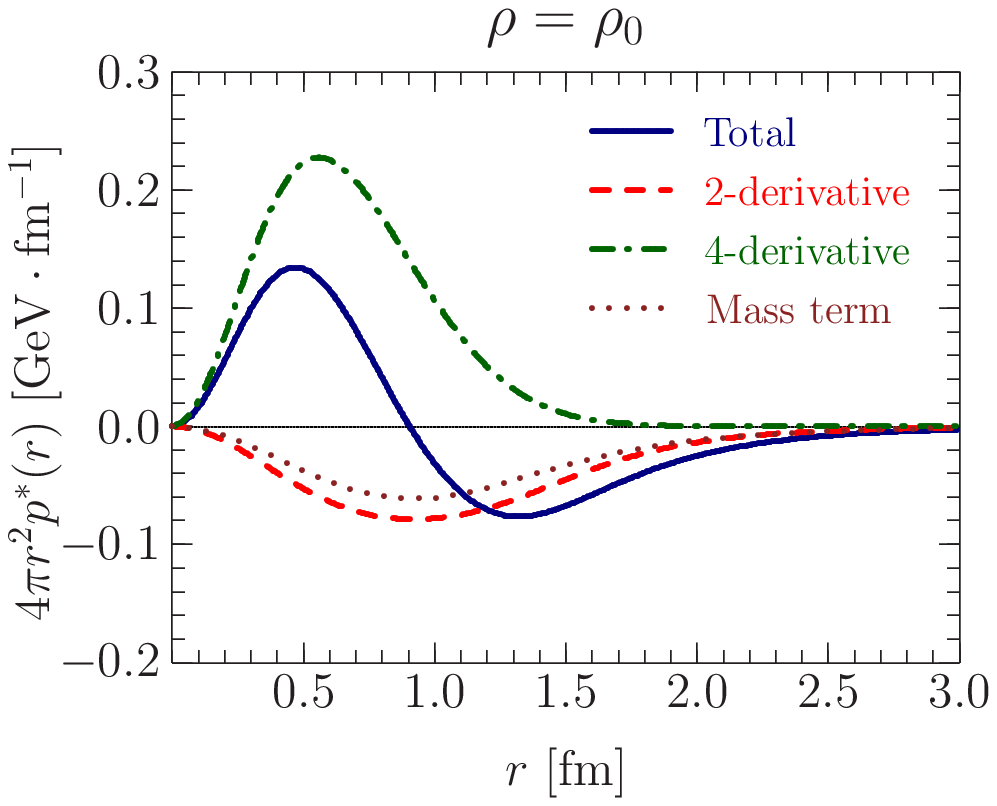}
}
\caption{(Color online)
The decomposition of the pressure densities $4\pi r^2 p^*(r)$ as
functions of $r$, in free space ($\rho=0$) and at $\rho=\rho_0$,
 in the left and right panels, respectively. The solid
curves denote the total pressure densities, the dashed ones represent
the contributions of the 2-derivative (kinetic) term, the
long-dashed ones are those of the 4-derivative (stabilizing) term, and
the dotted ones stand for those of the pion mass term.
}
\label{fig:3}
\end{figure}

\newpage \noindent
is right also in nuclear matter, which is explained in
Fig.~\ref{fig:3}. Comparing each contribution in the
right panel ($\rho=\rho_0$) to the corresponding one in the left
panel, we immediately realize that the 2- and 4-derivative terms are
suppressed as well as stretched to larger distance. On the other hand,
the mass term is enhanced at finite density. In order to
understand this feature more precisely, we examine the expression for
$p^*(r)$ given in Eq.(\ref{Eq:pressure}). The relevant coefficient in
the kinetic term is $F_{\pi,s}^*$ that becomes smaller as $\rho$
increases, so that its contribution to $p^*(r)$ gets diminished. In
the meanwhile, $e^*$ governs the strength of the contribution of the
stabilizing term. It grows as $\rho$ is turned on, which leads to
lessen the contribution of the stabilizing one to $p^*(r)$. In the
case of the pion mass term contribution to $p^*(r)$, the pertinent
coefficient is $m_\pi^{*2}F_{\pi,s}^{*2}=m_\pi^2 F_\pi^2\alpha_s$,
which becomes larger with $\rho$
increased. That results in the enhancement of the pion mass
term. While all these contributions undergo changes in nuclear
matter, the stability condition of Eq.~(\ref{eq:stable}) is still
satisfied. Integrating each contribution, we obtain for the case of
free space:
\begin{equation}
\int_0^\infty dr\, r^2 p(r) \;=\;  \left\{
\begin{array}{rl}
-9.550\, \mathrm{MeV}  & \mbox{  for the kinetic term} \\
12.446\, \mathrm{MeV}  & \mbox{  for the Skyrme term} \\
-2.896\, \mathrm{MeV}  & \mbox{  for the pion mass term}\,,
\end{array}\right.
\end{equation}
whereas for the case of $\rho=\rho_0$
\begin{equation}
\int_0^\infty dr\, r^2 p^*(r) \;=\;  \left\{
\begin{array}{rl}
-7.741\, \mathrm{MeV}  & \mbox{  for the kinetic term} \\
13.652\, \mathrm{MeV}  & \mbox{  for the Skyrme term} \\
-5.911\, \mathrm{MeV}  & \mbox{  for the pion mass term} \,.
\end{array}\right.
\end{equation}
We see that adding up all contributions becomes zero,
as required by the stability condition (\ref{eq:stable}).

\begin{figure}[t]
\centerline{
\includegraphics[height=4.62cm]{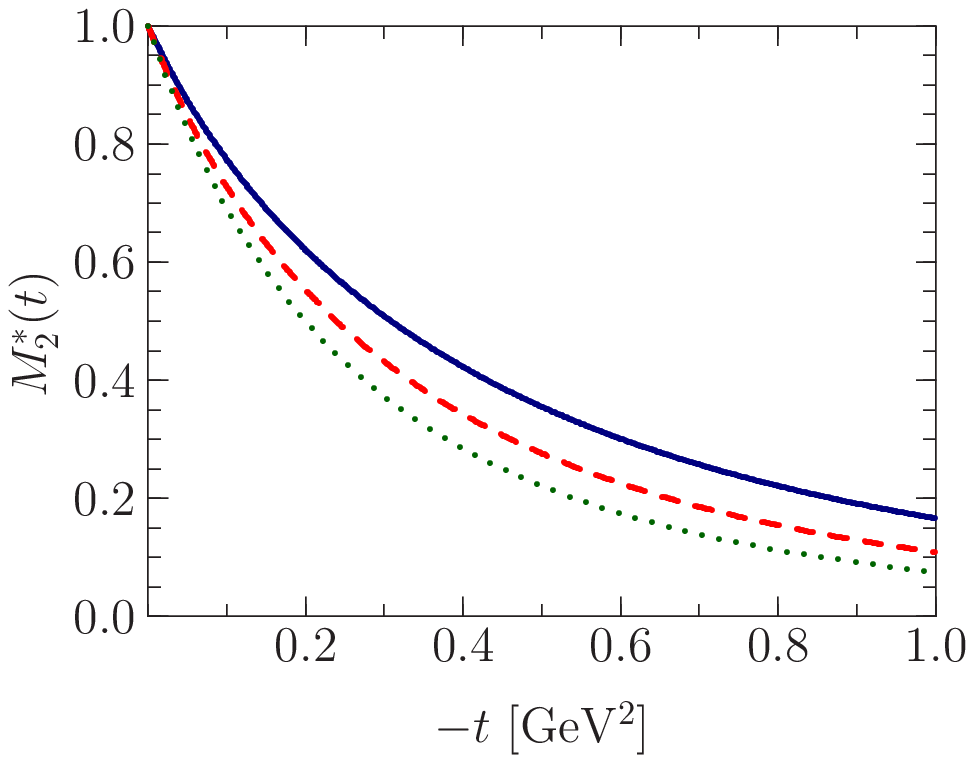}
\includegraphics[height=4.62cm]{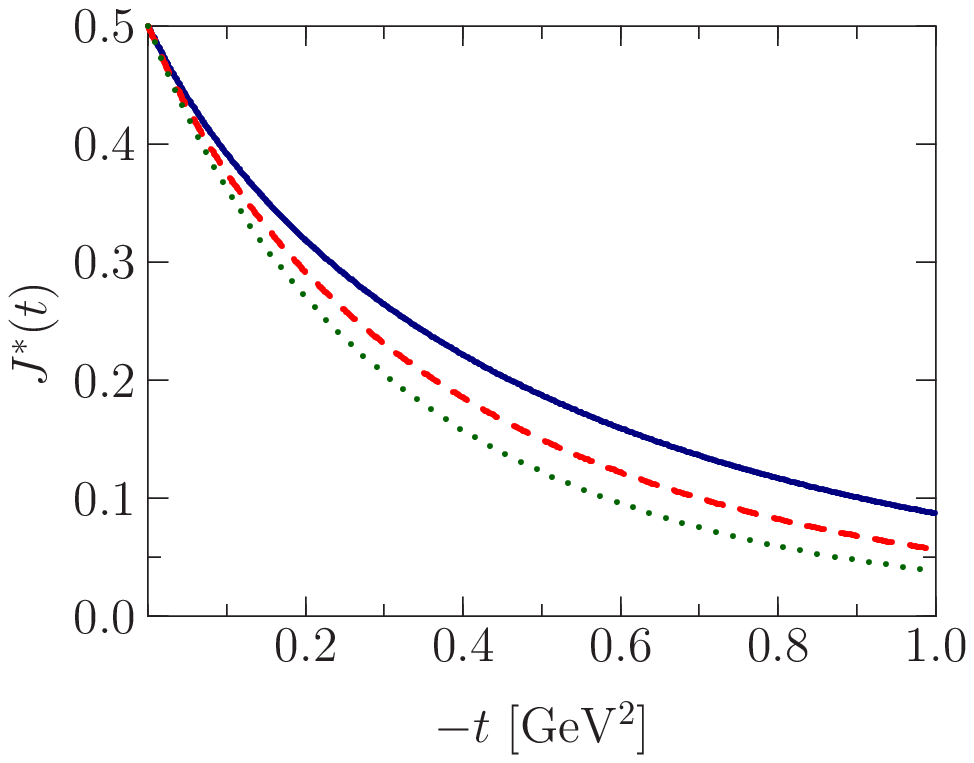}
\includegraphics[height=4.62cm]{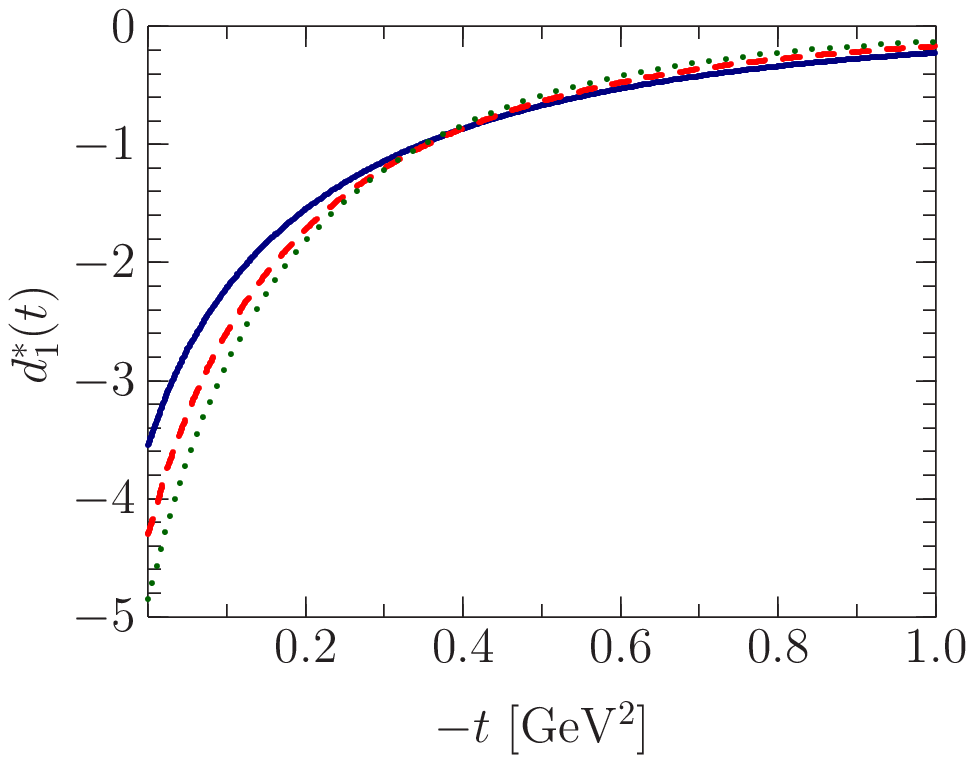}
}
    \caption{(Color online)
The EMT form factors of the nucleon $M_2^*(t)$, $J^*(t)$, and
$d_1^*(t)$ as functions of $t$ for the nucleon. Notations are the same
as in Fig.~\ref{fig:1}.
}
\label{fig:4}
\end{figure}
Having performed the Fourier transforms of the densities
$T_{00}^*(r)$, $\rho_J^*(r)$, and $p^*(r)$ discussed so far, we
immediately obtain the nucleon EMT form factors. Figure~\ref{fig:4}
draws the resulting EMT form factors as functions of $t$. The
form factors $M_2^*(t)$ and $J^*(t)$ fall off more rapidly as $\rho$
increases. On the other hand, $d_1^*(t)$ is rather distinguished from
the other two form factors. While $M_2^*(t)$ and $J^*(t)$ are
constrained to be $1$ and $1/2$ at $t=0$ as shown in
Eq.~(\ref{eq:norm}), $d_1^*(t)$  does not have such a constraint.
However, had one scaled $d_1^*(t)$ so that their values
may coincide at $t=0$, $d_1^*(t)$ would have ended up with almost
the same response to $\rho$ as $M_2^*(t)$ and $J^*(t)$.
As shown in Fig.~\ref{fig:4}, all the EMT form factors get stiffer
near the zero momentum transfer, as the density of nuclear matter
increases. It implies that all the corresponding radii get enlarged with
$\rho$ increased, as was already discussed for the case of $\langle
r_{00}^2\rangle^*$ and $\langle r_{J}^2\rangle^*$.

All EMT form factors can be well approximated by dipole-type
parametrizations $F^*(t) = F^*(0)/(1-t/M_{F^*}^{*2})^2$
with the generic dipole masses $M_{F^*}$ listed in Table~\ref{tab:2}.
\begin{table}[h]
\caption{\label{tab:2}
The dipole masses of the EMT form factors $M_2^*(t)$, $J^*(t)$ and
$d_1^*(t)$ which approximate well the results in Fig.~\ref{fig:4}.
}
    \begin{ruledtabular}
    \begin{tabular}{cccc}
$\rho/\rho_0$&
$M_{M_2^*}^*$ [GeV] &
$M_{J^*}^*$ [GeV]   &
$M_{d_1^*}^*$ [GeV] \\
    \hline
0    & 0.840 & 0.856 & 0.592 \\
0.5 & 0.756 & 0.798 & 0.582 \\
1.0 & 0.702 & 0.760 & 0.578 \\
\end{tabular}
\end{ruledtabular}
\end{table}

\newpage
\textbf{5.}
So far we have considered the medium effects in the range
$0 \le \rho \le \rho_0$ where our approach based on linear
response theory of pions in medium~\cite{Ericsonbook} is applicable.
The description in terms of hadronic degrees of freedom is expected
to become inappropriate and to break down at densities beyond~$\rho_0$.
It is interesting to remark that by continuing our in-medium modified
Skyrme model to $\rho > \rho_0$ we find indications for this expected
breakdown. We would like to stress that we do not expect the naive
extrapolation of the linear response
in Eqs.~(\ref{Eq:alpha_p},~\ref{Eq:alpha_s}) to work quantitatively.
However, the results are qualitatively  insightful.

\begin{figure}[h]
\centerline{
  \includegraphics[height=6.6cm]{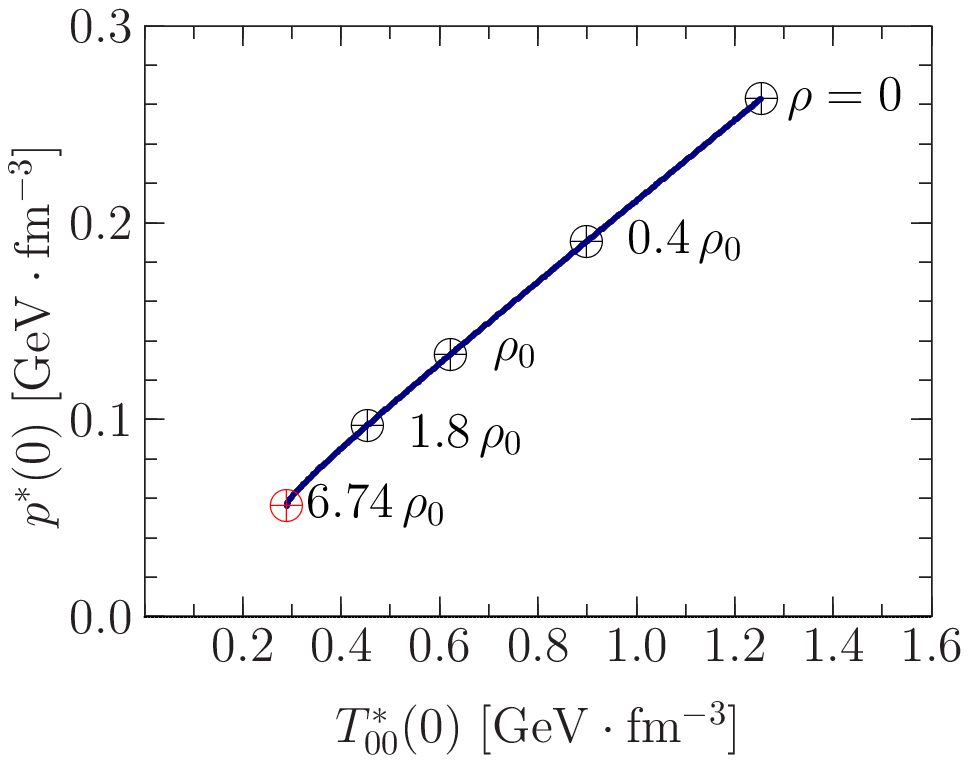}\hspace{0.5cm}
  \includegraphics[height=6.6cm]{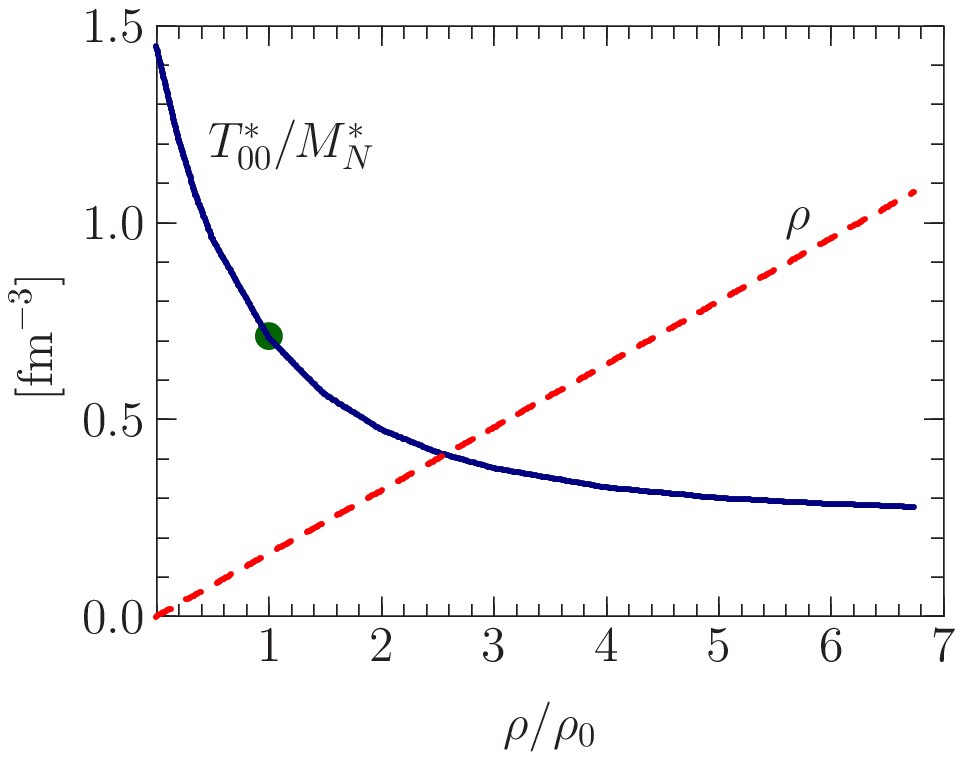}
}
\caption{(Color online) In the left panel, the correlated change of
  $p^*(0)$ and $T_{00}^*(0)$ are drawn with $\rho$ varied. In the right
  panel, the $T_{00}^*/M_N^*$ and $\rho$ depicted as a function of
  $\rho/\rho_0$.  The maximal density is given as about $6.74\,\rho_0$,
  above which the Skyrmion does not exist anymore. The filled circle
  on the solid curve represents the value of $T_{00}^*/M_N^*$ at
  normal nuclear matter density.
}
\label{fig:5}
\end{figure}

The left panel of Fig.~\ref{fig:5} shows the correlated modifications
of the pressure and energy densities in the center of the nucleon. We
already have seen from Figs.~\ref{fig:1} and \ref{fig:2}, both
$p^*(0)$ and $T_{00}^*(0)$ decrease as $\rho$ grows.
This trend continues also at higher
densities, and the interesting point is that
$p^*(0)$ and $T_{00}^*(0)$ are diminished at the same rate.
This is natural if we consider nuclear matter density as an
external parameter. The response of a stable system to variations
of external parameters must be such that the pressure increases
if the energy density does. The relation of $p^*(0)$ and
$T_{00}^*(0)$ remains linear to a good approximation, until the Skyrmion
ceases to exist at the ``maximal nuclear density''
$\rho_{\rm max}=6.74\,\rho_0$. This maximal value arises because
$\alpha_p(\rho)\to 0$ for $\rho\to\rho_{\rm max}$. In this limit
the energy functional (\ref{Eq:Esol}) has no minimum. This can
be shown by means of the Derrick theorem \cite{Derrick}, which is
equivalent to the stability condition (\ref{eq:stable}) \cite{Cebulla:2007ei}.
In other words, in the limit $\rho\to\rho_{\rm max}$ the stability
condition breaks down.
We stress that this maximal density
appears only in the mathematical description of
our medium-modified Skyrmion,
and has nothing to do with the physical critical point
relevant to the restoration of chiral symmetry.

The fact that the Skyrmion ceases to exist as $\rho\to\rho_{\rm max}$
is of interest from a mathematical point of view. Of far
greater relevance is that physical description of the nucleon in
medium breaks down before this point is reached.
This is because at $\rho\approx(2\mbox{--}3)\rho_0$ two things happen.
Firstly, the normalized energy density $T_{00}^*(0)/M_N^*$
(alternatively one could also consider the baryon number density)
becomes comparable with the density of the surrounding nuclear
medium (see the right panel of Fig.~\ref{fig:5}). Secondly, the size
of the nucleon, as measured for instance in terms of the square root
of the mean square radii or $r_0^*$, becomes comparable to
$\rho^{-1/3}$ which implies a spatial overlap of the nucleons in the
nuclear medium. At this point it makes no sense anymore to mention
an ``individual nucleon'' in nuclear medium, and a description
in terms of more appropriate degrees of freedom becomes
necessary (``quark matter'').
It is very interesting to observe that the medium-modified
Skyrme model signals in this way its limitations.

\vspace{0.8cm}
\textbf{6.}
In the present work, we investigated the energy-momentum
tensor form factors of the nucleon in nuclear matter, based on the
in-medium modified Skyrme model. Employing all the parameters fixed
in previous works, we derived the densities relevant to the energy,
the angular-momentum, the pressure, and the shear-force distributions
in the nucleon. We first examined how the energy and angular-momentum
densities were modified in nuclear matter. In general their values in
the center of the nucleon are suppressed but they are
stretched to larger distance in nuclear matter.
This was documented by the increase of the corresponding
mean square radii with nuclear matter density.

We also analyzed
the medium modifications of the pressure density. Again, its value in
the center of the nucleon decreases as nuclear matter density
grows. The change of the pressure density showed a similar behavior to
the energy
density in matter. The position where the pressure changes
sign increases with nuclear matter density increased, and the
pressure density falls off in medium more slowly than that in free
space. Since the pressure plays an essential role in the stability of
the Skyrmion, we scrutinized each contribution of the kinetic,
stabilizing, and mass terms.
With nuclear matter density increased, the contributions of the
kinetic and stabilizing terms are suppressed in magnitude, while that
of the mass term becomes larger, and all contributions 
extend to larger distances. The stability condition was shown to be
satisfied also in nuclear matter. Moreover in nuclear matter the
constant $d_1^*$ is negative as a consequence of stability.
In particular, we observe that medium effects further decrease
the value of $d_1^*$.

The energy-momentum tensor form factors $M_2(t)$ and $J(t)$, which are 
constrained to be 1 and 1/2 at zero momentum transfer, were
found to fall off more rapidly as nuclear matter density increased.
The absolute magnitude of the values of the
$d_1^*(t)$ form factor increased in medium, compared to that in
free space. The medium effects on its $t$-dependence were
more or less similar to the other two form factors.
Our results may pave a way towards a better understanding of
the discrepancy between predictions obtained from different models
of nuclei \cite{Polyakov:2002yz, Guzey:2005ba, Liuti:2005gi}.

The present results for the modification of the EMT form factors and
related properties, with nuclear density varied from zero to normal
nuclear matter density and beyond that, indicate a breakdown of the
description of the nucleon in nuclear matter densities at
$\rho\approx(2\mbox{--}3)\rho_0$. At such high densities the nucleon
in nuclear matter starts to loose its identity, and it would finally
melt away, which implies that a description in different degrees of
freedom becomes appropriate at such high densities.

\begin{acknowledgments}
The works of H.-Ch.~K. and U.~Y are supported by Basic Science
Research Program through the National Research Foundation of Korea
(NRF) funded by the Ministry of Education, Science and Technology
(Grant Number: 2012004024 (H.-Ch.~K.)  and Grant Number:
2011-0023478 (U.~Y.)), respectively. P.~S.\ is supported by DOE
contract No. DE-AC05-06OR23177, under which Jefferson Science
Associates, LLC operates Jefferson Lab.
\end{acknowledgments}


\end{document}